\documentclass[%
reprint,
superscriptaddress,
nofootinbib,
amsmath,amssymb,
aps,
pra,
]{revtex4-2}

\usepackage[normalem]{ulem} 
\usepackage{url}
\usepackage{hyperref} 

 \usepackage{textcomp}
 \newcommand{\textapproxx}{\raisebox{0.5ex}{\texttildelow}}

\usepackage[utf8]{inputenc}
\usepackage{mathtools}
\usepackage{algorithm2e}
\usepackage[caption=false]{subfig}
\usepackage{physics}
\usepackage{graphicx}
\usepackage{dcolumn}
\usepackage{bm}
\usepackage{verbatim}
\usepackage{xcolor}

\usepackage{amsmath}

\DeclareMathOperator*{\argmin}{arg\,min}

\definecolor{forestgreen}{rgb}{0.13, 0.55, 0.13}

\newcommand{\rn}{\bm{\Tilde{r}}}
\newcommand{\ri}{\bm{r}}

\newcommand{\rhon}{\Tilde{\rho}}
\newcommand{\rhoi}{\rho}

\newcommand{\rp}{\bm{\hat{r}}}

\begin{document}

\title{Quantum State Reconstruction in a Noisy Environment via Deep Learning}

\author{Angela Rosy Morgillo{\href{https://orcid.org/0009-0006-6142-0692}{\includegraphics[scale=0.004]{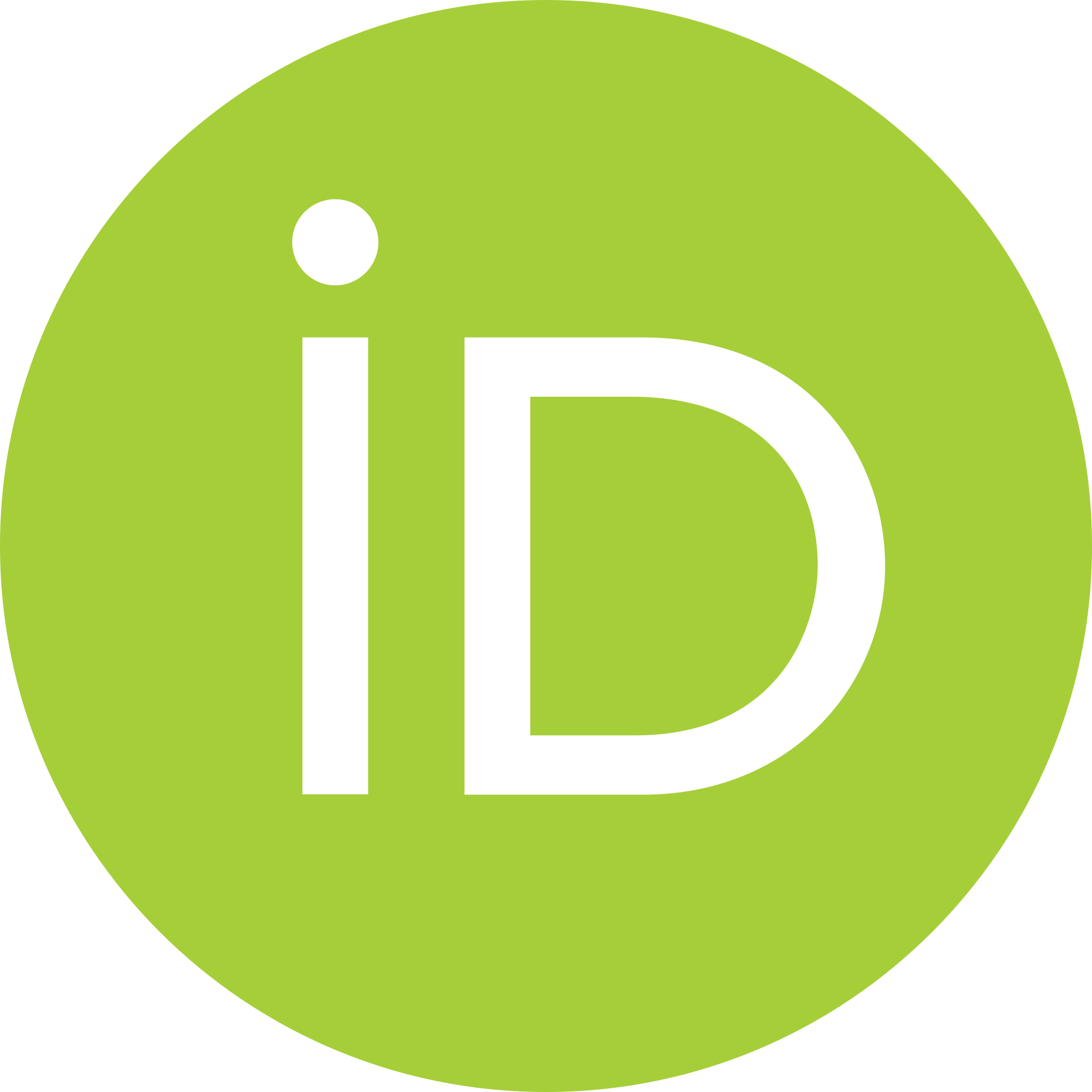}}}}
\email{angelarosy.morgillo01@universitadipavia.it}
\affiliation{Dipartimento di Fisica, Università di Pavia, Via Bassi 6, I-27100, Pavia, Italy}
\affiliation{INFN Sezione di Pavia, Via Bassi 6, I-27100, Pavia, Italy}

\author{Stefano Mangini{\href{https://orcid.org/0000-0002-0056-0660}{\includegraphics[scale=0.004]{orcid.png}}}}
\thanks{Current address:  ${}^1$\textit{Algorithmiq Ltd, Kanavakatu 3C 00160 Helsinki, Finland}. ${}^2$\textit{QTF Centre of Excellence, Department of Physics, University of Helsinki, P.O. Box 43, FI-00014 Helsinki, Finland.}}
\affiliation{Dipartimento di Fisica, Università di Pavia, Via Bassi 6, I-27100, Pavia, Italy}
\affiliation{INFN Sezione di Pavia, Via Bassi 6, I-27100, Pavia, Italy}

\author{Marco Piastra{\href{https://orcid.org/0000-0003-2556-5254}{\includegraphics[scale=0.004]{orcid.png}}}}
\affiliation{Dipartimento di Ingegneria Industriale e dell'Informazione, Università di Pavia, via Ferrata 5, I-27100, Pavia, Italy}

\author{Chiara Macchiavello{\href{https://orcid.org/0000-0002-2955-8759}{\includegraphics[scale=0.004]{orcid.png}}}}
\affiliation{Dipartimento di Fisica, Università di Pavia, Via Bassi 6, I-27100, Pavia, Italy}
\affiliation{INFN Sezione di Pavia, Via Bassi 6, I-27100, Pavia, Italy}

\date{\today}

\begin{abstract}
Quantum noise is currently limiting efficient quantum information processing and computation. In this work, we consider the tasks of reconstructing and classifying quantum states corrupted by the action of an unknown noisy channel using classical feedforward neural networks. By framing reconstruction as a regression problem, we show how such an approach can be used to recover with fidelities exceeding $99\%$ the noiseless density matrices of quantum states of up to three qubits undergoing noisy evolution, and we test its performance with both single-qubit (bit-flip, phase-flip, depolarising, and amplitude damping) and two-qubit quantum channels (correlated amplitude damping). Moreover, we also consider the task of distinguishing between different quantum noisy channels, and show how a neural network-based classifier is able to solve such a classification problem with perfect accuracy.
\end{abstract}

\maketitle

\section{\label{sec:level1}INTRODUCTION}
One of the main problems in quantum information processing and computation is that quantum systems can be corrupted by unwanted interactions with the environment. Therefore, the incorporation of robust quantum error correction and mitigation strategies is of paramount importance to realize the full potential of quantum information processing.

Despite the effectiveness of quantum error correction protocols in preserving information, they often require significant overhead and resources. Quantum error mitigation techniques, on the other hand, focus on reducing the impact of noise without fully correcting it, making them more feasible for near-term quantum devices~\cite{cai2022quantum, kandala2019error}. Examples are readout mitigation techniques to correct measurement errors~\cite{denBergModelFreeReadout2022, AlistairReaoutMitig2021, bravyi2021mitigating}, noise deconvolution methods to retrieve ideal expectation values of generic observables evaluated on a system subject to a known noise before measurement~\cite{mangini2021qubit, roncallo2023multiqubit}, probabilistic error cancellation~\cite{van_den_berg_probabilistic_2023} and data driven approaches such as zero-noise extrapolation~\cite{IBMkimEvidence2023} and Clifford data regression~\cite{Czarnik2021errormitigation, giurgica2020digital, lowe2021unified} to mitigate noise happening during a quantum computation.

Another area of great interest is deep learning, which has achieved impressive successes over the past years, with generative pre-trained large language models now leading the way~\cite{AttentionisAllyouNeed2023, Brown2020languageGPT3, deng2022benefits}. Deep learning models have excelled in diverse areas, from image and speech recognition models~\cite{he2016deep, kamath2019deep} to playing games~\cite{DeepMindMuZero2020}, reaching and often surpassing human-level performances. These advancements highlight the vast potential of deep learning in revolutionizing numerous fields, including quantum computation and information.

Indeed, deep learning techniques have shown great promises also for quantum information processing applications, as they were leveraged successfully in, e.g., experimental phase estimation tasks~\cite{lumino2018experimental}, automating the development of QCVV (Quantum characterization, validation and verification) protocols~\cite{scholten2019classifying}, learning quantum hardware-specific noise models~\cite{zlokapa2020deep}, increasing measurement precision of quantum observables with neural networks~\cite{torlai2020precise}, quantum error mitigation~\cite{kim2022quantum, kim2020quantum, sack23}, identifying quantum protocols such as teleportation or entanglement purification~\cite{wallnofer2020machine}, classification and reconstruction of optical quantum states~\cite{ahmed2021classification}, and quantum state estimation~\cite{lohani2020machine}. 

\begin{figure*}[!ht]
\label{fig:basic}
\centering
\subfloat[Quantum state reconstruction]
{\includegraphics[width=.8\textwidth]{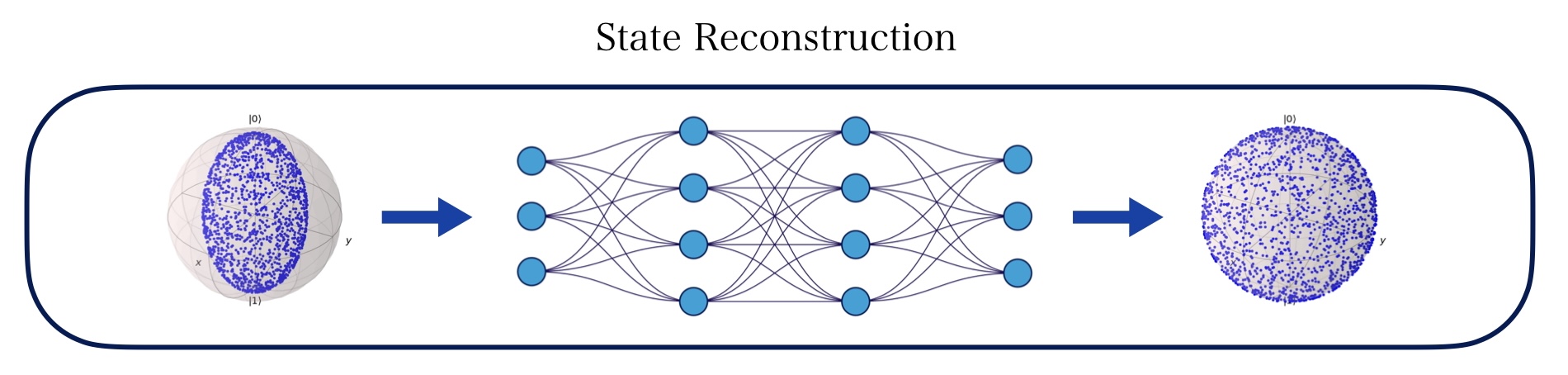}}\\
\subfloat[Quantum noise classification]
{\includegraphics[width=.8\textwidth]{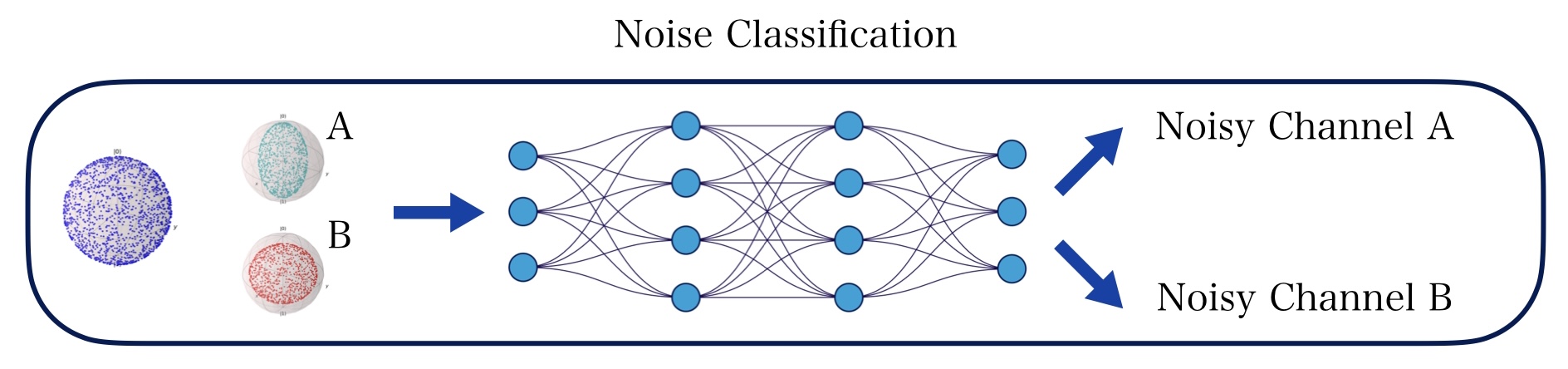}} 
\caption{Outline of the neural network-based noise reconstruction and classification protocols. \textbf{(a)} Noisy Bloch vectors, representing quantum states affected by noise, serve as input to a feed-forward neural network. The quantum state reconstruction protocol aims to recover the original noiseless quantum states from the observed noisy Bloch vectors, utilizing the neural network model. \textbf{(b)} Both noisy and noiseless Bloch vectors are fed into the neural network as input. The network is specifically designed for a classification task, where the output provides a label representing the type of noise acting on the noiseless quantum state.}
\label{fig:minimaldataset}
\end{figure*}

In this work, we leverage machine learning techniques based on feed-forward neural networks to deal with the task of recovering noise-free quantum states when they undergo an undesired noise evolution. In fact, while it is well known that quantum noisy channels cannot be physically inverted in general, this may be achieved by means of classical post-processing methods~\cite{mangini2021qubit, roncallo2023multiqubit, van_den_berg_probabilistic_2023}. In particular, since neural networks are universal approximators~\cite{hornik1989multilayer}, they can be used to learn a mapping that effectively inverts that effect of noise, and hence they can be used to reconstruct noiseless quantum states. Specifically, let $\rn$ indicate the (generalised) Bloch components of a noisy quantum state, our goal is to train a neural network $h_{\bm{w}}(\cdot)$ to output the Bloch vector of the ideal noiseless state $\rn \rightarrow h_{\bm{w}}(\rn) = \ri$, where $\ri$ is the ideal Bloch vector of the state before it undergoes noise process. We explore several combinations of single- and two-qubit noisy channels acting on systems of up to three qubits, and also study the effect of using different loss functions for training, and show that our neural network-based method can reach quantum state reconstruction fidelities higher than 99.9\%. The main idea of the proposed method is summarised in Figure~\ref{fig:basic}.

In addition to regression tasks, we also show how feed-forward neural networks can be used for classifying different quantum channels based on the effect they have on quantum states. In particular, using as inputs Bloch vectors $\qty[\rn, \ri]$ obtained with different channels, the network will output a label corresponding to the quantum channel that has been applied to $\ri$ in order to produce $\rn$. Also in this case, we achieve almost perfect channel classification accuracy.

The rest of the manuscript is organised as follows. In Section~\ref{sec:methods} we formally introduce the problem and the neural network used to obtain the quantum state reconstruction. In Section~\ref{sec:results} we present the results obtained for the reconstruction of pure and mixed states, and we introduce the noise classification problems that can be solved similarly with neural networks. In Section~\ref{sec:conclusion} we summarise all our results and possible improvements of our method.

\section{\label{sec:methods} Methods}
In this section, we formalise the quantum communication problem we want to tackle, that is, the recovery of noiseless quantum states undergoing an undesired noisy evolution. We first start by introducing the notation for describing an $n$-qubit quantum state in terms of its Bloch components, and then move on discussing the neural network approach used in this work, including details on the optimisation procedure and the construction of the training and test dataset.

\subsection{Reconstruction of noisy Bloch vectors}
The state of an $n$-qubit quantum system is described by its density matrix $\rho \in \mathbb{C}^{2n \times 2n}$, which can be expressed in the Pauli basis as follows~\cite{nielsen_chuang_2010, Ozols2007GeneralizedBV}
\begin{equation}
\label{eq:bloch_components}
    \rho = \frac{1}{2^n} \qty(\mathbb{I}_{2^n} + \bm{r} \cdot \bm{P})
\end{equation}
where $\bm{r} \in \mathbb{R}^{4^n-1}$ is the \textit{generalised Bloch vector}, and $\bm{P} = (P_1,\,\hdots,\, P_{4^n-1})$ is a vector containing the multi-qubit Pauli basis, obtained by considering tensor products of single-qubit Pauli matrices, that is $P_i = \sigma_1^{(i)} \otimes \cdots \otimes \sigma_n^{(i)}$, with $\sigma_k \in \{\mathbb{I}, X, Y, Z\}$.

Quantum channels are completely positive trace preserving (CPTP) maps whose action on a state $\rho$ can be expressed in Kraus form as~\cite{nielsen_chuang_2010}
\begin{equation}
\label{eq:noise_kraus}
    \rhoi \longrightarrow \rhon = \mathcal{N}(\rhoi) = \sum_{i} E_i\,\rho\,E_i^\dagger\,,
\end{equation}
where $\{E_i\}$ are the Kraus operators of the channel $\mathcal{N}(\cdot)$, satisfying the trace preserving condition $\sum_iE_i^{\dagger}E_i=\mathbb{I}$. In our experiments, we consider various single-qubit noisy channels (bit-flip $\mathcal{X}_p$, phase-flip $\mathcal{Z}_p$, bit-phase-flip $\mathcal{Y}_p$, general Pauli $\mathcal{P}_{\bm{p}}$, depolarizing $\mathcal{D}_p$ and amplitude damping $\mathcal{A}_{p\gamma}$), as well as a correlated two-qubit amplitude damping channel. We refer to Appendix~\ref{app:noisychan} for an extended discussion on the quantum noise models used in this work. 

Given a noisy channel $\mathcal{N}(\cdot)$, our goal is to obtain through a learning procedure an optimised neural network that receives noisy Bloch vectors $\{\rn_k\}$,
and outputs the corresponding noiseless vectors $\{\ri_k\}$. In other words, we are looking for the function $h(\cdot)$ which inverts the action of the noise on the Bloch components of the quantum states, namely
\begin{align}
\label{eq:summary_eq}
    &h(\rn)= \ri\,, \quad \text{with} \nonumber\\
    &\rhoi = \frac{1}{2^n} \qty(\mathbb{I}_{2^n} + \bm{\ri} \cdot \bm{P}) \\
    &\rhon = \mathcal{N}(\rhoi) = \frac{1}{2^n} \qty(\mathbb{I}_{2^n} + \bm{\rn} \cdot \bm{P})\nonumber\,.
\end{align}

\subsection{Reconstruction with neural networks}
We provide a concise overview of the fundamental aspects of neural networks, discussing their relevance in addressing the task of quantum state reconstruction.

\subsubsection{Generation of the training set}
The initial phase of addressing a regression problem involves constructing a valid dataset. In our specific case, the training (and validation) set consists of pairs of noisy and noiseless Bloch vectors
\begin{equation}
    \label{eq:training_set}
    \mathcal{T} = \qty{\qty(\rn_m,\, \ri_m)}_{m=1}^{M}\,,
\end{equation}
which are obtained by evolving some input quantum states $\rhoi_m$ through the noisy channel under investigation, thus obtaining the noisy states $\rhon_m$. The Bloch components of these density matrices are then computed as $r_i = \Tr[\rho\, P_i]$ for $i=1,\hdots, 4^n-1$~\eqref{eq:bloch_components}.

The choice of the (generalized) Bloch vector as element of the dataset is twofold: first, each quantum state is characterized by its own vector, which grants a unique representation of the state; and secondly a vector as input fits naturally in the processing structure of a feed-forward neural network. 

The input quantum states we consider are uniformly distributed in the space of quantum states. For the case of pure states $\rho_m = \dyad{\psi_m}$, these are obtained by sampling states $\ket{\psi_m}$ from the Haar distribution~\cite{edelman_rao_2005, meckes_2019}, while for uniformly distributed mixed states, these can be generated either starting from uniformly distributed pure states by means of an appropriate rescaling~\cite{HARMAN20102297, rubinstein2016simulation} or by using the Ginibre ensemble~\cite{random_mat, ginibre, forestbenchmarking}. 

The cardinality $|\mathcal{T}| = M$ of the dataset is contingent upon the specific problem under consideration and, as demonstrated in Sec.~\ref{sec:results}, has a direct impact on the performance of the network. As the quantum computational resources to generate the training set $\mathcal{T}$ may be demanding experimentally, one generally has to find a compromise between achieving high reconstruction accuracies, and the number of samples (i.e. quantum states) included in the dataset.

\subsubsection{Feed-forward neural networks}
In this work, we analyze data using deep feed-forward neural networks, which are parametric models that process information in layer-wise fashion through a repeated application of similar operations, as shown in Fig.~\ref{fig:basic}(a).

These neural networks consist of an \textit{input layer} responsible for data loading, followed by multiple \textit{hidden layers} to process the information, and finally, an \textit{output layer} to obtain the computation's result. Each layer consists of a set of individual nodes known as \textit{neurons}, and while input and output layers have a number of neurons matching the dimension of the input and the output respectively, the number of neurons in the hidden layers is an architectural hyperparameter to be chosen in advance by the designer. For example, the action of a feed-forward neural network with two hidden layers and trainable parameters $\bm{\theta}$, can be expressed as
\begin{equation}
\label{eq:ffnn}
\begin{split}
    \hat{\bm{y}} & = \text{NN}_{\bm{\theta}}(\bm{x})\\
    & = \bm{w} \cdot g\qty(\boldsymbol{W}^{[2]}~g\qty(\boldsymbol{W}^{[1]}\boldsymbol{x} + \boldsymbol{b}^{[1]}) + \boldsymbol{b}^{[2]}) + b
\end{split}
\end{equation}
where $\bm{x} \in \mathbb{R}^d$ and $\hat{\bm{y}} \in \mathbb{R}^{p}$ are the input and output vectors, $\boldsymbol{W}^{[1]} \in \mathbb{R}^{h_1 \times d}$ and $\bm{b}^{[1]} \in \mathbb{R}^{h_1}$ are the trainable parameters for the first hidden layer, $\boldsymbol{W}^{[2]} \in \mathbb{R}^{h_2 \times h_1}$ and $\bm{b}^{[2]}\in \mathbb{R}^{[h_2]}$ are the trainable parameters for the second hidden layer, $\bm{w} \in \mathbb{R}^{p}$ and $b \in \mathbb{R}$ are the trainable parameters for the output layer, and $g(\cdot)$ is a non-linear activation function which is applied element-wise to the entries of the vectors. As previously mentioned, $h_1$ and $h_2$ are hyperparameters that represent the number of hidden neurons for each respective layer.

In our simulations, we explore different architectures using networks with 2 or 3 hidden layers using $h_i \in \{64, 128\}$ hidden neurons per layer, while the input and output layers have dimension $d=p=4^n-1$, as they are employed to represent the components of the Bloch vectors. For the activation function, as customary in machine learning, we adopt the Rectified Linear Unit (ReLU), defined as $g(x) = \text{ReLU}(x) \coloneqq \max(0,x)$.

\subsubsection{Performance metrics}
Given the dataset and the trainable model, we discuss the figures of merit employed for training and evaluating the neural network's performance in quantum state reconstruction and noise classification tasks.
In the context of quantum state reconstruction, we have tested two possible alternatives coming from the classical and quantum information domain respectively, namely the Mean Squared Error (MSE) between the reconstructed and ideal Bloch vectors, and the quantum infidelity between the reconstructed and original quantum states. 

In noise classification problems, we used both categorical cross-entropy~\eqref{eq:cat} and accuracy metrics~\eqref{eq:acc} in order to assess how effectively the neural network can distinguish between different types of noisy channels.

\paragraph*{MSE.---} The mean squared error is the most common measure of performances for regression problems in machine learning and consists of the Euclidean distance between vectors, which in our case becomes
\begin{align}
    \label{eq:euclid_dist}
     \ell(\bm{\theta}, \bm{\ri}_i) = \norm{\bm{\ri}_i - \hat{\bm{r}}_i(\bm{\theta})}^2 = \norm{\bm{\ri}_i - \text{NN}_{\bm{\theta}}(\bm{\rn}_i)}^2,
\end{align}
where $\bm{\ri}_i$ is the noiseless Bloch vector (see Eq.~\eqref{eq:training_set}), and $\rp_i(\theta) =\text{NN}_{\bm{\theta}}(\rn_i)$ is the one predicted by the neural network, with trainable parameters $\bm{\theta}$, when receiving as input the noisy Bloch vector $\rn_i$. Then, the mean squared error function over the entire dataset $\mathcal{T}$ of size $|\mathcal{T}|=M$, is
\begin{equation}
\label{eq:mse_loss}
\begin{split}
     \mathcal{L}_\text{MSE}(\bm{\theta};\, \mathcal{T}) &= \frac{1}{M}\sum_{i=1}^{M}\ell(\bm{\theta}, \bm{\ri}_i)\\
     & =  \frac{1}{M}\sum_{i=1}^{M}\|\bm{\ri}_i - \hat{\bm{r}}_i\|^2 \quad (\bm{\rn_i}, \bm{\ri_i}) \in \mathcal{T}\,.
\end{split}
\end{equation}

\paragraph*{Infidelity.---} The quantum fidelity is a measure of distance for quantum states, and given the quantum nature of the data under investigation, it is particularly suited to assess the reconstruction performances of the neural network. Given two density matrices $\rho$ and $\sigma$, their fidelity is defined as~\cite{nielsen_chuang_2010}
\begin{equation}
    \label{eq:fid}
    F(\rho, \sigma) \coloneqq \Tr\qty[\sqrt{\sqrt{\rho}\,\sigma\,\sqrt{\rho}}]^2\,,
\end{equation}
with $0\leq F(\rho, \sigma)\leq 1$, where the second equality holds if and only if the states are equal $\rho = \sigma$. The \textit{infidelity} between two quantum states is then defined as $I(\rho, \sigma) \coloneqq 1 - F(\rho, \sigma)$. 

Despite being suited to measure the distance of quantum states, the complex functional dependence of the infidelity on the Bloch vectors of the density matrices ---hence the parameters of the neural network--- often leads to numerical instabilities when it is used as the loss function to drive the training of the neural network, eventually impairing the optimisation process. For this reason, when $\rho$ and $\sigma$ are pure states, we instead use directly the simplified but equivalent expression for the fidelity 
\begin{equation}
\label{eq:fid_pure}
    F(\rho,\sigma) = \Tr[\rho\,\sigma]\,,\quad \text{for } \rho, \sigma \text{ pure}\,,
\end{equation}
while if the states are mixed, we use an alternative measure of distance proposed in~\cite{wang2008alternative}
\begin{equation}
\label{eq:fidmix}
    F(\rho, \sigma) = \frac{\abs{\Tr[\rho\,\sigma]}}{\sqrt{\Tr[\rho^2] \Tr[\sigma^2]}}\,,\quad \text{for } \rho, \sigma \text{ mixed}\,.
\end{equation}
The fidelity in Eq.~\eqref{eq:fidmix} reaches the maximum 1 if and only if $\rho = \sigma$. However, it differs from the standard fidelity reported in Eq.~\eqref{eq:fid}, since it is not monotonic under quantum operations, meaning it is neither concave nor convex, and it includes a normalization factor, resulting in a scaled fidelity measure, when one of the two states is pure, i.e. if $\sigma=\dyad{\psi}$, then $F(\rho, \dyad{\psi}) = \langle \psi |\rho | \psi \rangle / \Tr\rho^2$.

From the fidelity expressions in Eqs.~\eqref{eq:fid_pure} and \eqref{eq:fidmix}, we calculate the corresponding infidelity and perform an average over the entire dataset, yielding
\begin{equation}
\label{eq:inf_loss_pure}
\begin{split}
     \mathcal{L}_\text{INF}(\bm{\theta};\, \mathcal{T}) &= \frac{1}{M}\sum_{i=1}^{M}1 -F(\rho_i, \sigma_i),
\end{split}
\end{equation}
where $\rho_i$ and $\sigma_i$ are the density matrices computed respectively from the Bloch vectors $\bm{\ri}_i$ and $\hat{\bm{r}}_i = \rp_i(\theta) =\text{NN}_{\bm{\theta}}(\rn_i)$, with $(\bm{\rn_i}, \bm{\ri_i}) \in \mathcal{T}$. 

Moreover, it is worth noticing that for single-qubit density matrices the fidelity~\eqref{eq:fid} can be further simplified and expressed in terms of the Bloch vectors as~\cite{fidjo}
\begin{equation}
\label{eq:fidvec}
\begin{split}
    F(\bm{r}, \bm{s}) = \frac{1}{2} \qty(1 + \bm{r} \cdot \bm{s} + \sqrt{(1 - \|\bm{r}\|^2) (1 - \|\bm{s}\|^2)} ),
\end{split}
\end{equation}
where $\bm{r}, \bm{s} \in \mathbb{R}^3$ are the Bloch vectors for $\rho$ and $\sigma$ respectively. Specifically, for the particular case when both states are pure $\norm{\bm{r}}^2=\norm{\bm{s}}^2 = 1$, then the expression for the infidelity corresponds to the mean squared error up to a prefactor
\begin{equation}
\label{eq:fidmse}
    I(\bm{r}, \bm{s}) = 1 - \frac{1}{2}(1+\bm{r}\cdot\bm{s}) 
      = \frac{\norm{\bm{s}-\bm{r}}^2}{4}\,.
\end{equation}
We refer to Appendix~\ref{fidformula} for more details on the derivation of Eqs.\eqref{eq:fidvec} and~\eqref{eq:fidmse}. Notably, to the best of our knowledge, apart for single-qubit states, there is no straightforward connection between the fidelity of quantum states and the Euclidean distance of their generalised Bloch vectors. 

Finally, in order to assess the performance of the optimised neural network we introduce the Average Test Fidelity (ATF), defined as the mean fidelity between the predicted quantum states and their corresponding ideal counterparts averaged over a test dataset $\mathcal{\tilde{T}}$ which was not used during training. The ATF is calculated as
\begin{equation}
    \label{eq:atf}
    \text{ATF}(\bm{\theta}; \mathcal{\Tilde{T}}) = \frac{1}{N}\sum_{i=1}^{N} F(\rho_i, \sigma_i),
\end{equation}
where $N$ is the cardinality of the test set, $F(\cdot,\,\cdot)$ is the fidelity in Eq.~\eqref{eq:fid}, and $\rho_i$ and $\sigma_i$ are the density matrices computed respectively from the Bloch vectors $\ri_i$ and $\rp_i=\text{NN}_{\bm{\theta}}(\rn_i)$, with $(\bm{\rn}_i, \bm{\ri}_i) \in \mathcal{\tilde{T}}$. A high average test fidelity, typically exceeding 99.9\%, indicates that our neural network is capable of accurately reconstructing the corrupted quantum states. 

\paragraph*{Categorical Cross-Entropy.---} The categorical cross-entropy is one of the most common measures to evaluate the performance of classification models~\cite{goodfellow2016deep}. Given a classification task with $C$ different classes, the categorical cross-entropy quantifies the disparity between the predicted probability distribution of the classes and the true class labels, and it is mathematically defined as
\begin{equation}
\label{eq:cat}
    \text{CCE} \coloneqq -\sum_{i=1}^{N} \sum_{j=1}^{C} y_{ij} \log(p_{ij}),
\end{equation}
where $M$ is the number of samples to be classified, $y_{i} = (0,0, \hdots, 1_{c(i)}, \hdots, 0) \in \{0,1\}^C$ is the true probability distribution for the samples indicating that the $i$-th sample belongs to class $c(i)$, and $p_{ij} \in [0,\,1]$ is the model's predicted probability distribution for the $i$-th sample to belong to the $j$-th class. This metric penalizes the deviation between predicted and actual distributions, with lower values indicating better alignment between the model's predictions and the true labels.

\paragraph*{Accuracy.---} For classification tasks the evaluation of performance commonly revolves around the metric of accuracy, which quantifies the effectiveness of a model in correctly categorizing samples within a dataset. Mathematically, the accuracy is defined as the ratio of the number of correctly classified samples to the total number of samples, namely
\begin{equation}
\label{eq:acc}
\text{ACC} \coloneqq \frac{\text{number of correct predictions}}{\text{total number of predictions}},
\end{equation}
where a value of 1 indicates a perfect classification.

\subsubsection{Optimization}
Training the neural network means solving the minimisation problem
\begin{equation}
    \bm{\theta}_{\text{opt}} = \argmin_{\bm{\theta}} \mathcal{L}(\bm{\theta};\, \mathcal{T}),
\end{equation}
where $\bm{\theta}$ are the trainable parameters of the neural network, $\mathcal{T}$ is the training dataset as defined in Eq.~\eqref{eq:training_set}, and $\mathcal{L}$ is the loss function driving the learning process, which, as discussed in the previous section, in our case is either the mean squared error~\eqref{eq:mse_loss} or the average infidelity~\eqref{eq:inf_loss_pure}.

We optimize the neural networks using Adam, a variant of a stochastic gradient descent with adaptive estimation of first and second order moments~\cite{kingma2014adam}. For the results shown in Sec.~\ref{sec:results}, we report the combination of training hyperparameters (number of training epochs, batch size, and learning rate) that attains the best performances.

\section{\label{sec:results}Results}
We now present the results obtained for quantum state reconstruction and quantum noise classification using the proposed neural network methods. We start discussing the performance of the neural network in reconstructing noise-free quantum states corrupted by various single- and two-qubit noisy channels, and then proceed showcasing the network's capability to classify different quantum noisy channels accurately. Our results demonstrate high-fidelity state reconstruction and robust channel classification, thus revealing the potential of machine learning techniques for quantum information processing and computation. 

All simulations performed in this work are run with Qiskit~\cite{Qiskit} and TensorFlow~\cite{tensorflow2015-whitepaper}. 

\subsection{Quantum State Reconstruction}
In order to explore the reconstruction capabilities of the neural network approach, we first start by studying its performances in the simpler case of learning noisy single-qubit states, and then move on to the more complex case of multi-qubit systems. In both scenarios, we see a clear dependence of the performances on the amount of available data to train the model (size of the training set), and provided that data is sufficient the network is always able to restore noiseless quantum states from their noisy counterparts. 

Whenever we consider the task of reconstructing initially pure states, an auxiliary \textit{normalization layer} is added at the end of the neural network~\eqref{eq:ffnn} so that the outputs always consists of (Bloch) vectors with unit norm (for single-qubit states), as desired for pure states. Such normalisation constraint enforces the generation of physically consistent output states, which in turns effectively constrains the value of the infidelity loss function to remain in the physical regime\footnote{The fidelity~\eqref{eq:fid} can be larger than $1$ if the Bloch vectors have norms exceeding one, which is not possible for density matrices of quantum systems. If no constraint on the Bloch vectors is present, then the training process will simply push the neural network to output states of larger and larger norm to maximise the overlap.}. It's worth noting that even when the MSE is employed as the loss function, the same normalization layer is used to ensure that the predicted states maintain their physical integrity and adhere to the desired constraints. The effect of the normalization layer is simply to rescale each output as follows
\begin{equation}
\label{eq:normalisation}
    \text{NN}_{\bm{\theta}}(\rn) = \hat{\bm{r}} \longrightarrow \hat{\bm{r}} / \norm{\hat{\bm{r}}}_2\,.
\end{equation}

In addition, whenever we consider the case of reconstructing already mixed states $\rho$ of given purity $\Tr[\rho^2]$, the output of the neural network is additionally rescaled with the purity of the initial mixed state, $\hat{\bm{r}} \rightarrow \sqrt{\Tr[\rho^2]} \hat{\bm{r}} $.

\paragraph*{Single-qubit states.---}
In Table~\ref{tab:oneq} we summarize the results of the reconstruction process for various noisy channels, using different loss functions to drive the training process, and using pure or mixed states as inputs to the neural network. In all cases we observe a very good average test fidelity (ATF)~\eqref{eq:atf} at the end of training exceeding 99.6\%, thus showing the effectiveness of the proposed approach for reconstructing noisy states. As we now see, equally good performances are also obtained in the more complex task of inverting noise in multi-qubit systems.

\begin{table}[t]
\begin{ruledtabular}
\begin{tabular}{ccc}
 Channel & ATF (MSE) & ATF (INF)\\ \hline \\[-2.5ex]
 \multicolumn{3}{ c}{$|\mathcal{T}|=30$ Pure States}\\[0.1ex]
 \hline
 $\mathcal{X}_p$ ($p=0.2$) & 0.998 & 0.997 \\
 $\mathcal{Z}_p$ ($p=0.2$) & 0.997 & 0.998 \\
 $\mathcal{Y}_p$ ($p=0.2$) & 0.998 & 0.998 \\
 $\mathcal{P}_{\bm{p}}$ ($p_0 = 0.7, p_{1,2,3} = 0.1$) & 0.998 & 0.998\\
 $\mathcal{D}_p$ ($p=0.3$) & 0.996 & 0.997\\
 $\mathcal{A}_{p\gamma}$ ($p=0.5, \gamma=0.3$) & 0.998 & 0.997\\
 \hline\\[-2.5ex]
 \multicolumn{3}{ c}{$|\mathcal{T}|=30$ Mixed States}\\[0.1ex]
 \hline
  $\mathcal{Z}_p$ ($p=0.2$) & 0.999 & 0.999 \\
\end{tabular}
\end{ruledtabular}
\caption{\label{tab:oneq} Reconstruction of single-qubit quantum states. Training with both the MSE~\eqref{eq:mse_loss} and infidelity~\eqref{eq:inf_loss_pure} as loss functions yields good average test fidelities (ATF) at the end of the optimisation process. Results are reported for different noisy channels (bit-flip $\mathcal{X}_p$, phase-flip $\mathcal{Z}_p$, bit-phase flip $\mathcal{Y}_p$, general Pauli $\mathcal{P}_{\bm{p}}$, depolarizing $\mathcal{D}_p$, and general amplitude damping channels $\mathcal{A}_{p\gamma}$), and for both pure and mixed initial states.}
\end{table}

\paragraph*{Multi-qubit states.---} 
\begin{table}[!ht]
\begin{ruledtabular}
\begin{tabular}{ccc}
 Channel & ATF (MSE) & ATF (INF)\\ \hline\\[-2.5ex]
 \multicolumn{3}{ c}{$|\mathcal{T}| = 300$ Two-qubit States}\\[0.1ex]
 \hline
 $\mathcal{Z}_p \otimes \mathcal{I}$ ($p=0.2$) & 0.994 & 0.993 \\
 $\mathcal{Z}_p \otimes \mathcal{Z}_p$ ($p=0.2$)& 0.994 & 0.995 \\
 $\mathcal{Z}_p \otimes \mathcal{X}_p$  ($p=0.2$)& 0.994 & 0.995 \\
 $\mathcal{C}_{\text{AD}}$ ($\eta = 0.1$, $\mu = 0.2$) & 0.995 & 0.999\\
 \hline\\[-2.5ex]
 \multicolumn{3}{ c}{$|\mathcal{T}|=900$ Three-qubit States}\\[0.1ex]
 \hline
 $\mathcal{X}_p \otimes \mathcal{Z}_p \otimes \mathcal{Y}_p$ ($p=0.2$) & 0.999 & 0.999\\
\end{tabular}
\end{ruledtabular}
\caption{\label{tab:23} Results of pure multi-qubit states reconstruction using MSE and infidelity as loss functions. For two-qubit states, we studied scenarios involving: a phase-flip channel on the first qubit $\mathcal{Z}_p \otimes \mathcal{I}$, phase-flip channel on both qubits $\mathcal{Z}_p \otimes \mathcal{Z}_p$, phase-flip and bit-flip channels respectively on the first and second qubit $\mathcal{Z}_p \otimes \mathcal{X}_p$, and correlated amplitude damping $\mathcal{C}_{\text{AD}}(\eta,\, \mu)$. For three-qubit states, we considered the scenario characterized by bit-, phase-, and bit-phase-flip channels applied distinctly to all three qubits $\mathcal{X}_p \otimes \mathcal{Z}_p \otimes \mathcal{Y}_p$.}
\end{table}
We tested the reconstruction procedure also on systems of $n=2$ and $n=3$ qubit systems undergoing several noisy evolutions with both uncorrelated and correlated noisy channels, and we summarize the results in Table~\ref{tab:23}. For two-qubit systems, we considered the following noise maps: (\textit{i}) phase-flip channel with $p=0.2$ applied to the first qubit and nothing on the second indicated as $\mathcal{Z}_p \otimes \mathcal{I}$; (\textit{ii}) a phase-flip channel applied to both qubits with $p=0.2$, denoted as $\mathcal{Z}_p \otimes \mathcal{Z}_p$; (\textit{iii}) phase-flip channel on the first qubit and a bit-flip channel on the second qubit, both with $p=0.2$, denoted as $\mathcal{Z}_p \otimes \mathcal{X}_p$; \textit{(iv)} and finally we also have investigated a scenario where a correlated two-qubit amplitude damping channel $\mathcal{C}_{AD}$ ($\eta = 0.1$, $\mu = 0.2$), was applied to the system (see Eq.\eqref{eq:correlated_ad} in Appendix~\ref{app:noisychan} for a definition of the channel). 

For $n=3$ qubit systems instead, we tested the reconstruction performance with states subject to the composite channel $\mathcal{X}_p \otimes \mathcal{Z}_p \otimes \mathcal{Y}_p$. In this configuration, each qubit experiences a distinct quantum channel, including a bit-flip, phase-flip, and bit-phase-flip channel with a common noise parameter of $p=0.2$.

The results reported in Table~\ref{tab:23} again reveal a successful reconstruction of ideal density matrices through the use of a relatively simple feed-forward neural network, and thus confirm the effectiveness of the proposed method. As already mentioned previously, we stress again that to ensure the production of pure quantum states as output, a normalization layer has been incorporated. Specifically, by appropriately multiplying the normalisation layer in~\eqref{eq:normalisation}, the norm of the output Bloch vectors is constrained to $\sqrt{3}$ for two-qubit states, while it is set to $\sqrt{7}$~\cite{avron2020elementary} for three-qubit states. Comparing the cardinality of the training set $\abs{\mathcal{T}}$ used for the simulations in Tab.~\ref{tab:oneq} and Tab~\ref{tab:23}, we see that more samples in the training dataset are generally needed to ensure a good reconstruction for larger system sizes, as one would expect. A discussion on the impact of the available information on the reconstruction performances is the topic of the next section.

\begin{figure}[!ht]
    \centering
    \includegraphics[width=.45\textwidth]{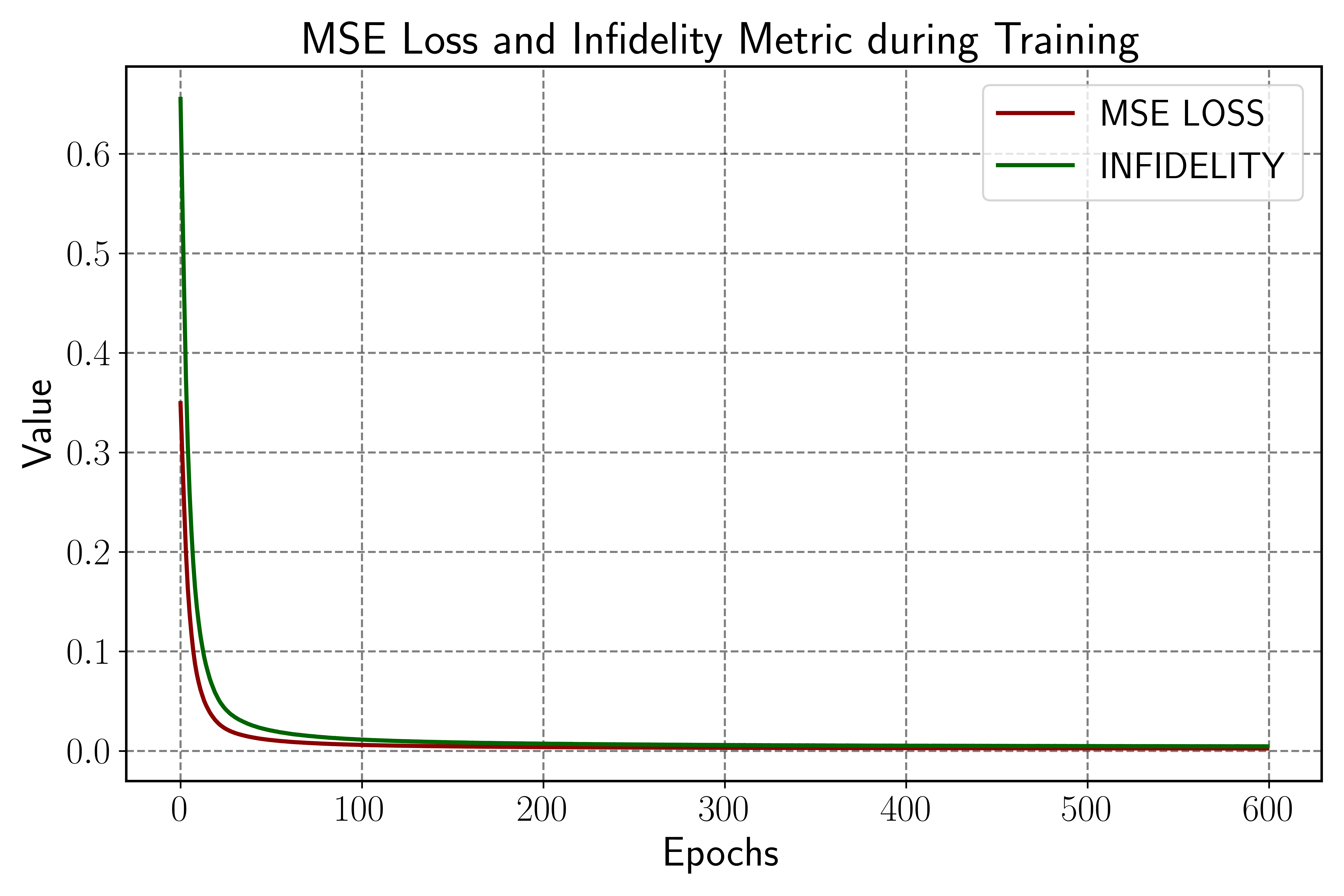}
    \caption{Optimisation process of the neural network to reconstruct a two-qubit state under application of the noisy channel $\mathcal{Z}_p \otimes \mathcal{X}_p$, using the MSE~\eqref{eq:mse_loss} and the infidelity~\eqref{eq:inf_loss_pure} as loss functions. Both metrics display a similar minimisation behaviour.}
    \label{fig:loss}
\end{figure}

In Fig~\ref{fig:loss} we report the evolution of the MSE~\eqref{eq:mse_loss} and infidelity~\eqref{eq:inf_loss_pure} loss functions during training for the case of $\mathcal{Z}_p \otimes \mathcal{X}_p$ applied to a two-qubit system. As clear from the picture, the optimisation process of both cost functions is straightforward, and interestingly they follow a similar minimisation behaviour, despite there is not a trivial relation between the two, as instead it happens for single qubit states, see Eq.~\ref{eq:fidmse}.

We conclude by noticing that, taking into account that an increase in the number of qubits leads to a proportional increase in the dataset cardinality and the neural network complexity, the reconstruction approach proposed in this work can be straightforwardly applied to any $n$-qubit quantum state.

\subsubsection{Impact of the dataset size on the reconstruction performances} 
\begin{figure*}[!ht]
\centering
\subfloat[]
{\includegraphics[width=.5\textwidth]{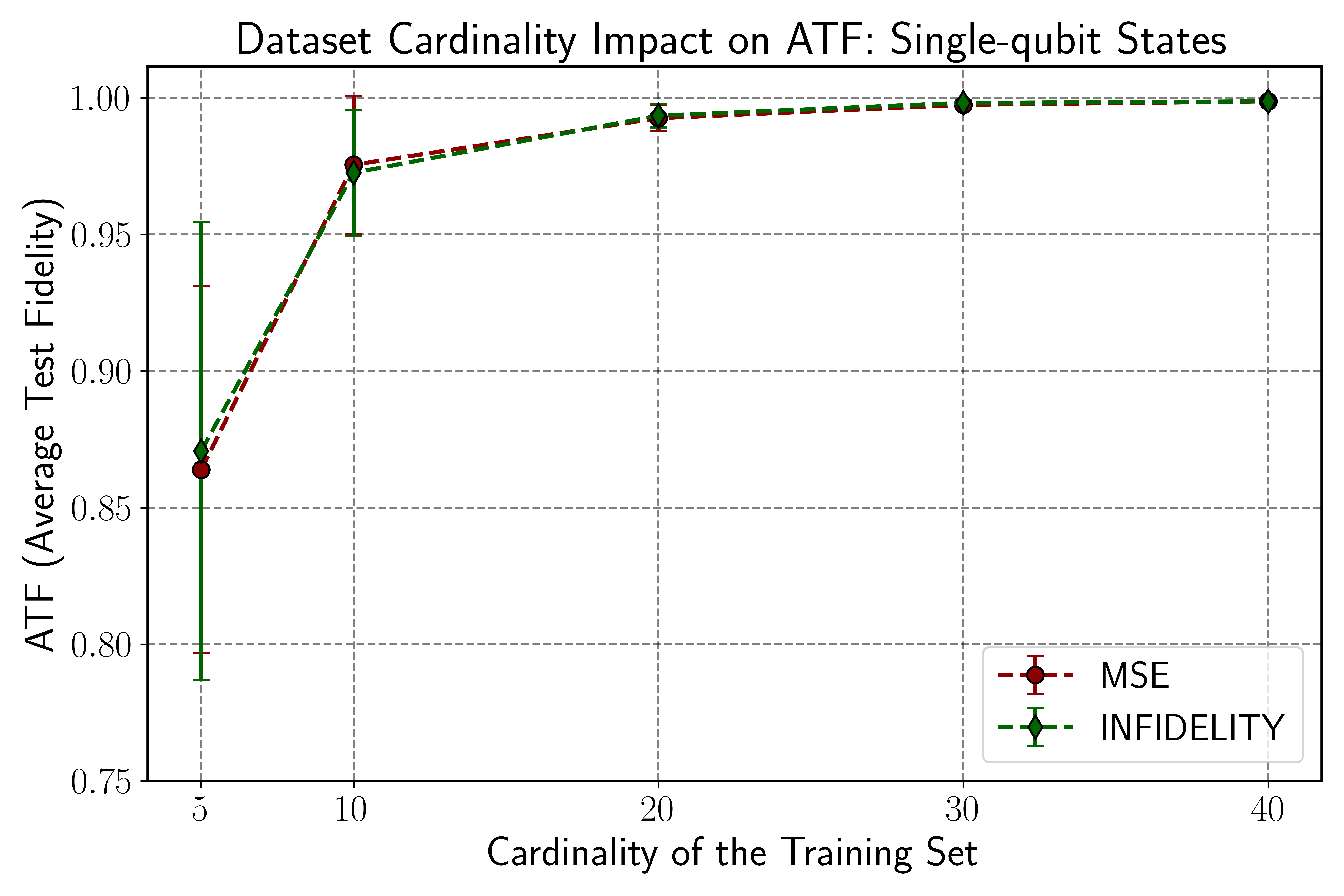}\label{subfig:a}}
\subfloat[]
{\includegraphics[width=.5\textwidth]{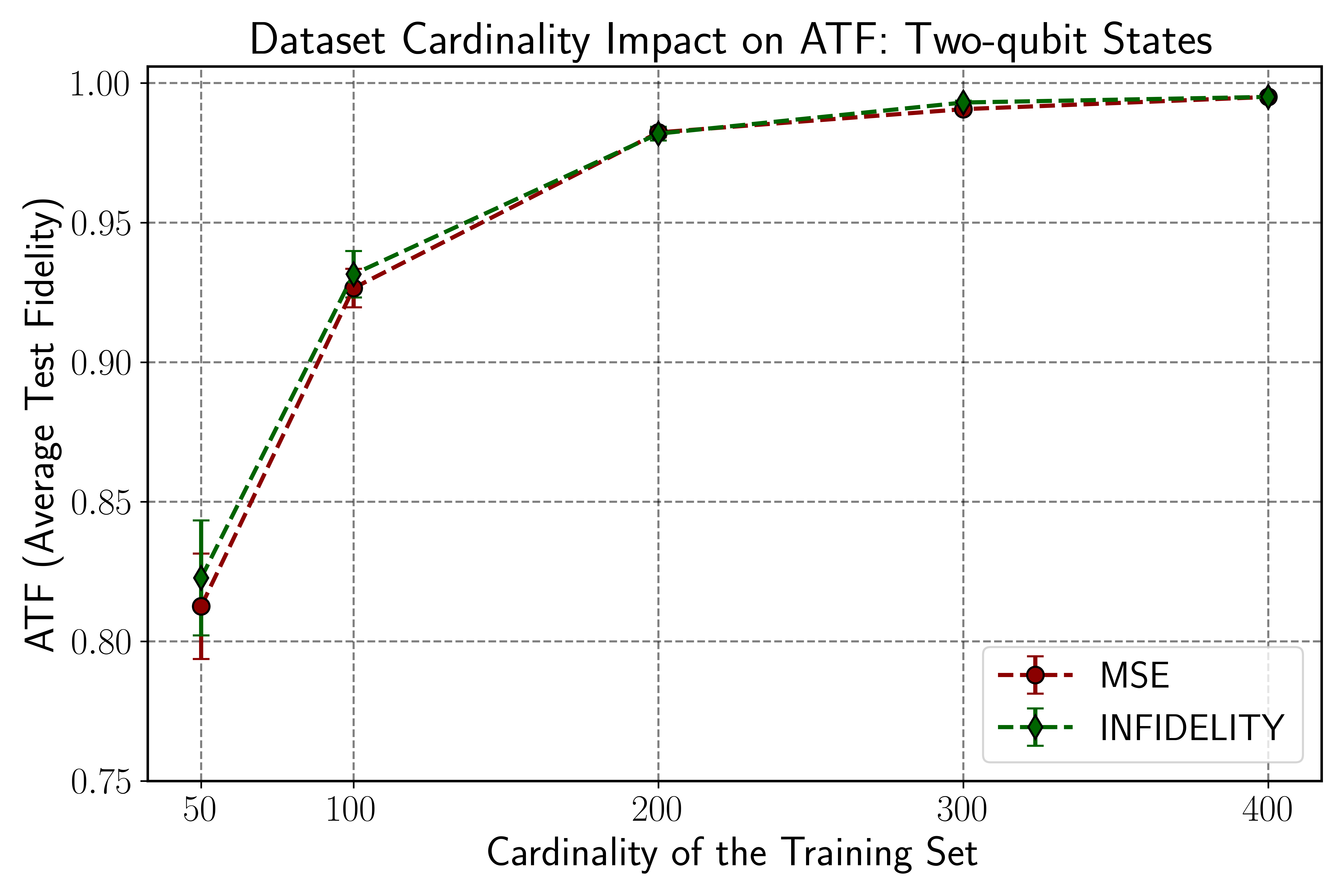}\label{subfig:b}}
\caption{Average Test Fidelity (ATF) obtained at the end of training when optimising the neural networks with training sets of different cardinality. For each cardinality, we repeat the training process 5 times using different initialisation of the parameters and different training data, and report the mean value and the standard deviation of the resulting ATFs. (a) Reconstruction of single-qubit states undergoing a phase-flip channel $\mathcal{Z}_{p}(p=0.2)$. (b) Reconstruction of two-qubit states undergoing the uncorrelated phase-flip channel $\mathcal{Z}_{p} \otimes \mathcal{Z}_{p}(p=0.2)$.}
\label{fig:md}
\end{figure*}

So far we have focused on assessing the reconstruction performances of the neural network assuming that enough data is available, and here instead analyse how such performances depends on the size of the training set. 

In Fig.~\ref{fig:md} we report the average test fidelity obtained at the end of training, for neural networks optimised using training sets of different cardinality. In panel~\ref{subfig:a} we show the data for the reconstruction of single-qubit states undergoing a phase-flip channel $\mathcal{Z}_{0.2}$, and in panel~\ref{subfig:b} the reconstruction of two-qubit states undergoing an uncorrelated two-qubit phase-flip channel $\mathcal{Z}_{0.2} \otimes \mathcal{Z}_{0.2}$.  

In both cases ---and for both the considered loss functions---, we observe that the use of a larger training set yields better reconstruction performances, up until a plateau is reached, but importantly also that satisfactory results can be achieved even with a limited number of samples. 

\subsection{Classification of noisy channels}
The application of neural networks to quantum information processing can be extended beyond quantum state reconstruction to that of classification of quantum channels. In particular, in this section we show how a neural network can be trained to discriminate between noisy channels based on the effect they have on some input states, a scenario which is graphically depicted in Fig.~\ref{fig:basic}. Our exploration encompasses a series of classification scenarios, including binary and multi-class classification.

In such classification problems, each data item in the training set is constructed by considering as input the noiseless Bloch vector $\ri_m$ appended to the noisy one $\rn_m$, and as output an integer label $y_m$ encoding which type of error was applied to $\ri_m$ to obtain $\rn_m$. Formally, given a classification task with $C$ possible classes, the training dataset is then defined as 
\begin{equation}
    \label{eq:classification_trainset}
    \mathcal{T}_{IN} = \big\{\big(\qty[\rn_m, \ri_m], y_m)\big)\big\}_{m=1}^M\,,
\end{equation}
where $\qty[\rn_m, \ri_m] \in \mathbb{R}^{2 \times 3}$ is the input to the neural network, and $y_m \in \{1,\hdots,C\}$ is the desired output. We refer to this training dataset with $\mathcal{T}_{IN}$, where the subscript indicates that we use as input the extended vector obtained by merging the \textit{ideal} and \textit{noisy} Bloch vector.

Furthermore, we consider the more complex scenario where every training data point comprises only the noisy vector $\rn_m$, accompanied with its associated noise label $y_m$. As the neural network is now provided with less information, this scenario presents a higher level of complexity and, as will become clear from the following results, it generally necessitates a larger corpus of data points. We refer to this type of training dataset with $\mathcal{T}_{N}$, where the subscript indicates that we use as input exclusively the \text{noisy} Bloch vector.

As standard with classification tasks in machine learning, for both scenarios we use a \textit{one-hot encoding} of the labels and train the neural networks using the categorical cross entropy~\eqref{eq:cat} as loss function~\cite{goodfellow2016deep}, and then measure the final performances with the accuracy metric~\eqref{eq:acc}.

\paragraph*{Binary classification.---}
We consider the binary classification problem ($C=2$) of discriminating single-qubit states subject either to phase-flip channel $\mathcal{Z}_p$, or amplitude damping channel $\mathcal{A}_{p\gamma}$. The training set is generated by sampling a number $\abs{\mathcal{T}}$ of random uniform single-qubit states and evolving half of them with the phase-flip channel, and the remaining half with the amplitude damping one. Classification accuracies at the end of the training procedure are reported in Table~\ref{tab:class}. 

\begin{table}[!t]
\caption{\label{tab:class} Results for quantum channel classification tasks, with corresponding accuracies~\eqref{eq:acc} obtained at the end of training and evaluated on test sets $\tilde{\mathcal{T}}$. We studied the binary classification task of distinguishing channels $\mathcal{Z}_p$ ($p=0.2$) vs. $\mathcal{A}_{p\gamma}$ ($p=0.5, \gamma=0.3$) in both variations ``IN" and ``N" for the training dataset~\eqref{eq:classification_trainset}, and the three-class classification problem for channels $\mathcal{Z}_p$ ($p=0.2$) vs. $\mathcal{A}_{p\gamma}$ ($p=0.5, \gamma=0.3$) vs. $\mathcal{D}_p$ ($p=0.3$) in the variant ``IN".}
\begin{ruledtabular}
\begin{tabular}{cc}
 Dataset cardinality & Accuracy (on test set $\Tilde{\mathcal{T}}$)\\ \hline\\[-2.5ex]
 \multicolumn{2}{ c}{Binary Classification} \\
 \multicolumn{2}{ c}{$\mathcal{Z}_p$ vs $\mathcal{A}_{p\gamma}$} \\
 \hline
 \raisebox{-0.5ex}{$|\mathcal{T}_{IN}| = 300$} & \raisebox{-0.5ex}{\phantom{0.}1\phantom{0} ($|\Tilde{\mathcal{T}}| = 100$)}\\
 $|\mathcal{T}_{N}| = 300$ & 0.92 ($|\Tilde{\mathcal{T}}| = 100$)\\
 \hline \\[-2.5ex]
 \multicolumn{2}{ c}{Ternary Classification}\\[-0.5ex] 
 \multicolumn{2}{ c}{$\mathcal{Z}_p$ vs $\mathcal{A}_{p\gamma}$ vs $\mathcal{D}_p$}\\[0.1ex] \hline
 \raisebox{-0.5ex}{$|\mathcal{T}_{IN}| = 960$} & \raisebox{-0.5ex}{\phantom{0.}{1}\phantom{0} ($|\Tilde{\mathcal{T}}| = 120$)} \\[-0.5ex]
\end{tabular}
\end{ruledtabular}
\end{table}

Remarkably, with a training set containing $\abs{\mathcal{T}_{IN}}=300$ samples, our model exhibits very good classification performances, reaching perfect accuracy $\text{ACC}=1$. In this case, we noticed that even with a reduced dataset comprising just a few dozen samples, the model is able to achieve almost perfect accuracy, even though the learning process becomes perceptibly less stable. 

On the other hand, with the noisy training set $\mathcal{T}_N$ with $\abs{\mathcal{T}_N}=300$ samples, the best accuracy obtained was $\text{ACC}=0.92$. As expected, as the noisy dataset $\mathcal{T}_N$ contains only information about the noisy Bloch vectors (and a label for the noisy channel that created them), more data is needed to reach good classification performances. Indeed, higher accuracies can then be obtained by using larger training datasets: for example, an accuracy of $\text{ACC}=0.98$ can be achieved with a training set containing 800 samples. 

\begin{figure}[!ht]
    \centering
    \includegraphics[width=.3\textwidth]{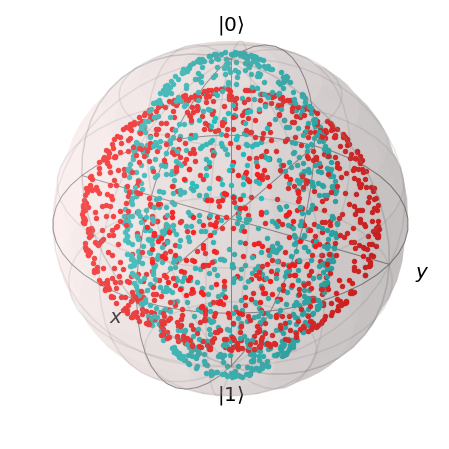}
    \caption{Deformed Bloch spheres obtained by applying a phase-flip channel $\mathcal{Z}_p(p=0.2)$ (light blue), and a generalised amplitude damping channel $\mathcal{A}_{p\gamma}(p=0.5,\, \gamma=0.3)$, to a set of uniformly distributed pure states. Note the non-trivial intersection between the two ellipsoids.}
    \label{fig:class}
\end{figure}

Note that when using dataset $\mathcal{T}_N$, learning the relation between noisy Bloch vectors belonging to the same channel implicitly translates to that of reconstructing the shape of the Bloch sphere generated by the two channels, which are reported in Fig.~\ref{fig:class}. However, as clear from the graphical representation, there is a nontrivial intersection between the two Bloch spheres, which implies that certain samples could reasonably be assigned to either classes. The imperfect accuracy and the general decrease in classification accuracy for the noise-only training dataset $\mathcal{T}_N$ shown in Table~\ref{tab:class}, can then be explained by such inherent ambiguity in the dataset which makes the classification task more difficult ---if not impossible--- to solve exactly.

\paragraph*{Multi-class classification.---} As a straightforward extension to the previous analysis, we also report results for the case of a multi-class classification task, where the network is asked to classify states generated by three different channels (phase-flip, amplitude damping and depolarizing channel), using a dataset of type $\mathcal{T}_{IN}$, that is containing both ideal and noisy Bloch vectors. We find again that the network is able to perfectly classify all the states, reaching a perfect final accuracy on a test set $\text{ACC}=1$, as reported in Table~\ref{tab:class}.

\section{\label{sec:conclusion}Conclusion}
In conclusion, our research underscores the remarkable effectiveness of deep neural networks in quantum information processing tasks, specifically in reconstructing and classifying quantum states undergoing unknown noisy evolutions. 

This study, as exemplified in our results, showcases the successful recovery of ideal (generalized) Bloch vectors with fidelities exceeding 0.99, even for quantum states involving up to three qubits, under different correlated and uncorrelated noisy channels, and using both classical (mean squared error) and quantum-inspired (infidelity) loss functions for training.

Furthermore, our investigation demonstrates the versatility of our neural network approach in classification problems, adeptly handling a wide range of noise patterns and consistently achieving remarkable classification accuracy across all test samples. Notably, in the context of discriminating between phase-flip and amplitude damping channels, our model achieves an outstanding classification accuracy of 98\%, highlighting its remarkable capacity to discern the relationships between states affected by similar noise sources, even when presented with the noisy vectors alone.


As we look ahead, an intriguing avenue for further exploration lies in examining the intricate connections between various fidelity measures~\cite{Liang_2019} used as training loss functions and their impact on the resulting test fidelities. This pursuit aims to identify the fidelity metric best suited to the specific characteristics of the problem at hand. Such investigations promise to advance our understanding of quantum information processing and open new horizons for practical applications in quantum technology.




\section{acknowledgements}
A.R.M. acknowledges support from the PNRR MUR Project PE0000023-NQSTI. C.M. acknowledges support from the National Research Centre for HPC, Big Data and Quantum Computing, PNNR MUR Project CN0000013-ICSC, and from the EU H2020 QuantERA ERA-NET Cofund in Quantum Technologies project QuICHE."
\appendix

\section{Noise Channels \label{app:noisychan}}
In this appendix we present the noise quantum channels used in the simulations to generate the datasets containing noisy Bloch vectors.

\paragraph{Bit-flip Channel.---} The bit-flip channel flips the qubit state from $|0\rangle$ to $|1\rangle$ and vice versa with probability $p$. Given the operator-sum representation in Eq.~\eqref{eq:noise_kraus}, its Kraus operators are
\begin{align}
    E_0 = \sqrt{1-p}\,\mathbb{I} && E_1 = \sqrt{p}\,X,\, \quad X = 
    \begin{pmatrix}
    0 & 1\\
    1 & 0\\
    \end{pmatrix}.
\end{align}

It is possible to represent the deformation that occurs on the Bloch sphere after the bit-flip noise occurred: the states on the $\hat{x}$ axis are left alone while the $\hat{y}-\hat{z}$ plane is uniformly contracted by a factor $2p$.

\paragraph{Phase-flip channel.---} This channel changes the sign of the component associated to the element of the computational basis $|1\rangle$ of the qubit state. The channel is represented by the operation elements
\begin{align}
    E_0 = \sqrt{1-p}\,\mathbb{I} && E_1 = \sqrt{p}\,Z,\, \quad Z = 
    \begin{pmatrix}
    1 & 0\\
    0 & -1\\
    \end{pmatrix}.
\end{align}
On the Bloch sphere, while the $\hat{x}-\hat{y}$ plane is contracted by a factor $2p$, the states on the $\hat{z}$ axis are left untouched.

\paragraph{Bit-phase-flip channel.---} This channel is a combination of a bit-flip and a phase-flip channel. Recalling that $Y=iXZ$, the Kraus operators of this channel are
\begin{align}
    E_0 = \sqrt{1-p}\,\mathbb{I} && E_1 = \sqrt{p}\,Y,\, \quad Y = 
    \begin{pmatrix}
    0 & -i\\
    i & 0\\
    \end{pmatrix}.
\end{align}
This noise acts on the Bloch sphere by leaving alone the states on the $\hat{y}$ axis and contracting the $\hat{x}-\hat{z}$ plane by a factor $2p$.

\paragraph{General Pauli channel.---} The general Pauli channel is a combination of bit-, phase- and bit-phase-flip channels, each one with its own intensity on the correspondent axis~\cite{mangini2021qubit}. In this case, the operators are
\begin{align}
  E_0 = \sqrt{p_0}\,\mathbb{I}  &&  E_1 = \sqrt{p_1}\,X, \\ E_2 = \sqrt{p_2}\,Y && E_3 = \sqrt{p_3}\,Z, 
\end{align}
with $p_1, p_2$ and $p_3$ probabilities of each error such that $p_0 + p_1 + p_2 + p_3 = 1$.

\paragraph{Depolarizing Channel.---} The depolarizing channel acts on a quantum state by leaving it untouched with probability $1-p$, or replacing it with the completely mixed state $\mathbb{I}/2$ with probability $p$. Its Kraus operators are
\begin{align}
    E_0 = \sqrt{1-\frac{3p}{4}}\,\mathbb{I} && E_1 = \frac{\sqrt{p}}{2}X \\ E_2 = \frac{\sqrt{p}}{2}Y && E_3 = \frac{\sqrt{p}}{2}Z\,.
\end{align}
In this case the entire Bloch sphere is uniformly contracted by a factor which depends on $p$. The channel can be generalized for $d$-dimensional ($d=2^n$ for $n$ qubits) quantum systems as
\begin{equation}
    \mathcal{E}(\rho) = (1-p)\rho + p\frac{\mathbb{I}_d}{d}\,.
\end{equation}

\paragraph{Generalized amplitude damping channel.---} This channel can be used to describe the energy dissipation in a quantum system to an environment at finite temperature~\cite{nielsen_chuang_2010}, and can be described with Kraus operators
\begin{equation}
\label{eq:E0}
    E_0 = \sqrt{p}\begin{pmatrix}
        1 & 0\\
        0 & \sqrt{1-\gamma}
    \end{pmatrix},
\end{equation}    
\begin{equation}
\label{eq:E1}
    E_1 = \sqrt{p}\begin{pmatrix}
        0 & \sqrt{\gamma}\\
        0 & 0
    \end{pmatrix}, 
\end{equation}
\begin{equation}
     E_2 = \sqrt{1-p}\begin{pmatrix}
        \sqrt{1-\gamma} & 0\\
        0 & 1
    \end{pmatrix}, 
\end{equation}  
\begin{equation}
    E_3 = \sqrt{1-p}\begin{pmatrix}
        0 & 0\\
        \sqrt{\gamma} & 0
    \end{pmatrix},
\end{equation}
where it is possible to define the stationary state
\begin{equation}
    \rho_{\infty} = \begin{pmatrix}
        p & 0 \\
        0 & 1-p
    \end{pmatrix}.
\end{equation}
This channel leads to a deformation of the Bloch sphere, where $\gamma$ regulates the shrinking of each component of the Bloch vector and $p$ indicates which is the fixed point.

\paragraph{Correlated amplitude damping channel.---} This is a two-qubit noise channel defined as the convex combination of two channels $\mathcal{N}_0$ and $\mathcal{N}_1$~\cite{d2013classical}
\begin{equation}
\label{eq:correlated_ad}
    \mathcal{N} (\rho) = (1-\mu)\mathcal{N}_0(\rho) + \mu \mathcal{N}_1(\rho),
\end{equation}
where $\mu \in [0,1]$ is a correlation parameter between the qubits, and $\mathcal{N}_0 = \sum_{j=0}^{3} A_j \rho A_j^\dagger$ and $\mathcal{N}_1 = \sum_{j=0}^{1} B_j \rho B_j^\dagger$ are noisy channels defined by Kraus operators $A_0 = E_0 \otimes E_0, A_1 = E_0 \otimes E_1, A_2 = E_1 \otimes E_0$ and $A_3 = E_1 \otimes E_1$ with $E_0$ and $E_1$ being the operators in Eqs.~\eqref{eq:E0} and~\eqref{eq:E1} with $p=1$, and 
\begin{equation}
    B_0 = \begin{pmatrix}
        1 & 0& 0& 0\\
        0 & 1 & 0& 0\\
        0 & 0 & 1 & 0\\
        0 & 0 & 0 & \sqrt{\gamma} \\ 
    \end{pmatrix}    ,
\end{equation}
\begin{equation}
     B_1 = \begin{pmatrix}
        0 & 0& 0& \sqrt{1 - \gamma}\\
        0 & 0 & 0& 0\\
        0 & 0 & 0 & 0\\
        0 & 0 & 0 & 0 \\
    \end{pmatrix}.
\end{equation}

\section{Fidelity of single qubit states}
\label{fidformula}
As proved in~\cite{fidjo}, the fidelity formula~\eqref{eq:fid} can be much simplified for one-qubit states and expressed in terms of Bloch vectors.
For an hermitian $2 \times 2$ matrix $M$ with positive eigenvalues, it holds
\begin{equation}
    \qty(\Tr \sqrt{M})^2 = \Tr[M] + 2 (\det M)^{1/2},
\end{equation}
and with $M = \sqrt{\rho}\,\sigma\,\sqrt{\rho}$ as in the definition of the fidelity~\eqref{eq:fid}, one obtains
\begin{equation}
    F(\rho, \sigma) = \Tr[\rho \sigma] + 2 (\det \rho \det \sigma)^{1/2}.
\end{equation}
Finally, expressing the density matrices in terms of their Bloch vectors, namely $\rho = (\mathbb{I} + \bm{r} \cdot \bm{P}) / 2$ and $\sigma = (\mathbb{I} + \bm{s} \cdot \bm{P}) / 2$ with $\bm{P} = (X, Y, Z)$, by explicit calculation one arrives at
\begin{equation}
    F(\bm{r}, \bm{s}) = \frac{1}{2} \qty(1 + \bm{r} \cdot \bm{s} + \sqrt{(1 - \|\bm{r}\|^2) (1 - \|\bm{s}\|^2)} )\,.
\end{equation}

With such expression at hand, it is then possible to prove that for pure single-qubit states minimizing the mean squared error loss is equivalent to minimizing infidelity. In fact, consider the Euclidean distance between the Bloch vectors
\begin{equation}
\label{eq:euclid}
    \ell(\bm{r}, \bm{s}) = \norm{\bm{r}-\bm{s}}^2 = \norm{\bm{r}}^2 + \norm{\bm{s}}^2 - 2\, \bm{r} \cdot \bm{s}\,,
\end{equation}
and the infidelity
\begin{equation}
    I(\bm{r}, \bm{s}) = 1 - \frac{1}{2} \qty(1 + \bm{r} \cdot \bm{s} + \sqrt{(1 - \|\bm{r}\|^2) (1 - \|\bm{s}\|^2)} )\,.
\end{equation}
Then, since we are computing the loss functions with Bloch vectors $\bm{r}$ and $\bm{s}$ of pure states ---the ideal noise-free Bloch vector and the predicted one by the neural network, see~\eqref{eq:training_set}---, it holds that $\norm{\bm{r}} = \norm{\bm{s}} = 1$, thus obtaining
\begin{equation}
    I(\bm{r}, \bm{s}) = 1 - \frac{1}{2}(1 + \bm{r} \cdot \bm{s}) \longrightarrow \bm{r} \cdot \bm{s} = -2\, I(\bm{r}, \bm{s}) + 1\,,
\end{equation}
and substituting in Eq.~\eqref{eq:euclid} one finally arrives at
\begin{equation}
    \ell(\bm{r}, \bm{s}) = 4\, I(\bm{r}, \bm{s}).
\end{equation}

\bibliography{bibliography}

\end{document}